\documentclass[11pt,a4paper,english]{article}
\usepackage{jcappub}

\usepackage{amsgen,amsmath,amstext,amsbsy,amsopn,amssymb}

\usepackage{graphicx}
\usepackage{epsfig}
\usepackage{multirow}
\usepackage{longtable}
\usepackage{hyperref}
\usepackage{color}
\usepackage{natbib}
\begin{document}

\title{Lagrangian bias in the local bias model}
\author[a,b]{Noemi Frusciante} 
\affiliation[a]{SISSA, International School for Advanced Studies,\\ 
             Via Bonomea  265, 34136, Trieste, Italy} 
\affiliation[b]{INFN, Sez. Trieste, \\
Via Valerio 2, 34127, Trieste, Italy}
\author[c,d]{and Ravi K. Sheth} 
\affiliation[c]{Center for Particle Cosmology, University of Pennsylvania,\\ 
             209 S 33rd Street,  Philadelphia, PA 19104, USA}
\affiliation[d]{Abdus Salam International Centre for Theoretical Physics, \\
 Strada Costiera 11, 34151, Trieste, Italy}

\emailAdd{nfruscia@sissa.it}
\emailAdd{sheth@ictp.it}
\date{\today}         

\abstract{
It is often assumed that the halo-patch fluctuation field can be written 
as a Taylor series in the initial Lagrangian dark matter density 
fluctuation field.  We show that if this Lagrangian bias is local, and 
the initial conditions are Gaussian, then the two-point cross-correlation 
between halos and mass should be linearly proportional to the mass-mass 
auto-correlation function.  
This statement is exact and valid on all scales; there are no 
higher order contributions, e.g., from terms proportional to products 
or convolutions of two-point functions, which one might have thought 
would appear upon truncating the Taylor series of the halo bias 
function.  In addition, the auto-correlation function of locally 
biased tracers can be written as a Taylor series in the 
auto-correlation function of the mass; there are no terms involving, 
e.g., derivatives or convolutions.  
Moreover, although the leading order coefficient, the linear bias 
factor of the auto-correlation function is just the square of that 
for the cross-correlation, it is the same as that obtained from 
expanding the mean number of halos as a function of the local density 
only in the large-scale limit.  In principle, these relations allow 
simple tests of whether or not halo bias is indeed local in 
Lagrangian space.  We discuss why things are more complicated in
practice.  We also discuss our results in light of recent work on the 
renormalizability of halo bias, demonstrating that it is better to 
renormalize than not.  We use the Lognormal model to illustrate many 
of our findings.}  

\arxivnumber{1208.0229}

\maketitle

\flushbottom

\section{Introduction}

Galaxies are biased tracers of the dark matter distribution \cite{nk1984, bbks1986}.  In the local bias model, the abundance of the biased tracers at a given position is assumed to be related to the mass at the same position.  The simplest version of this model, in which the smoothed overdensity field of the biased tracers $\delta_b$ is treated as though it were a deterministic function of the (similarly smoothed) real-space mass overdensity field $\delta_m$, has been the subject of much study \cite{as1988, pc1993, sw1998}.  Following ref.~\cite{fg1993}, it has become common to write the local model as
\begin{equation}
 \delta_b = f(\delta_m) = \sum_{i>0} \frac{b_i}{i!} \,(\delta_m^i - \langle\delta_m^i\rangle),
 \label{lb}
\end{equation}
where $b_i$ is the bias coefficient of order $i$.  Note that this model ensures $\langle\delta_b\rangle = 0$ by subtracting-off the $\langle\delta_m^i\rangle$ terms.\\  
One of the points we make in what follows is that one should, instead, normalize using a multiplicative factor which ensures that $\langle 1+\delta_b\rangle = 1$.  I.e., given an expansion of the form~(\ref{lb}), one should always work with 
\begin{equation}
 \delta_{B} \equiv \frac{1+\delta_{b} - \langle 1+\delta_b\rangle}
                        {\langle 1+\delta_b\rangle}.
  \label{lbcorrect}
\end{equation}

Although the model was invoked to describe the bias with respect to the late-time nonlinear Eulerian field $\delta_m^E$, since it is only invoked on large scales, it is often assumed, and sometimes explicitly used, to describe the bias with respect to the initial Lagrangian field $\delta_m^L$ (on large scales, these are expected to be similar).   However, the two best studied models of bias with respect to the Lagrangian field -- those associated with peaks, and patches which form halos, in a Gaussian random field -- behave rather differently from naive expectations based on the local bias model.  In particular, in the local bias model calculation of the cross-correlation between the biased tracers and the initial field, $\langle\delta_b\delta_m\rangle$, one proceeds by writing the bias function as a Taylor series, and then expanding order by order in $\delta_m$.  This means that one expects higher-order terms to contribute.  However, for peaks and in Gaussian initial conditions, the exact answer is 
 $\langle\delta_b\delta_m\rangle = b_1\,\langle\delta_m^2\rangle$
\cite{dcss2010}.  Therefore, in the expansion referred to above, all the higher-order terms must cancel out.  We show below that this is also true for another popular choice of the bias -- a Lognormal mapping between the biased field and $\delta_m$ -- and then that this is generically true in models where the bias is local and deterministic with respect to the initial Gaussian random field.  I.e., the cross-correlation is always only linearly proportional to the auto-correlation signal of the dark matter, although, in general, the linear bias factor need not equal $b_1$ of the Taylor series.  We then show that the auto-correlation function of the biased tracers can always be written as a Taylor series in the auto-correlation function of the original (unbiased) mass fluctuation field; derivatives and convolutions do not enter.  Along the way, our analysis connects to recent work on renormalized bias \cite{pm2006}, showing, e.g., that this renormalization is required if the Lognormal mapping is to make sensible predictions.  

We have structured the discussion as follows.  
In section~\ref{ln} we discuss the Lognormal mapping, since it turns out that all quantities of interest can be computed exactly (no truncation of the sums is required).  These exact expressions exhibit some curious properties which are not obvious if one trucates the sums.  Then we show that how one normalizes such purely formal expansions plays an important role, and why eq.~(\ref{lbcorrect}) is to be preferred over eq.~(\ref{lb}).  
Section~\ref{pks} explores a more complex example in which some of this simplicity is lost, before showing the general result in section~\ref{general}.  
Section~\ref{renorm} shows how to proceed if the full expansion is not available, so one is constrained to work with a truncated series.  Section~\ref{hermiteBias} connects this analysis to some of the earliest work on this subject (ref.~\cite{as1988}), discussing how the coefficients of the Taylor series expansion of $\delta_B$ in terms of $\delta_m$ are related to those obtained from expanding $\xi_{BB}$ in terms of $\xi_{mm}$.  Section~\ref{halos} discusses halo bias in the context of these results, and section~\ref{porq} revisits a technical point about halo bias, first made by ref.~\cite{sl1999} but often overlooked, which complicates the use of cross-correlations for testing the hypothesis that Lagrangian halo bias is indeed local, a subject of much recent interest \cite{mg2011, psp2012, css2012, mps2012}.  A final section summarizes.  

\section{Local Lagrangian bias}\label{llb}

\subsection{The Lognormal model and the usual expansion}\label{ln}
We begin with the usual model, eq.~(\ref{lb}), with $\delta_m=\delta_L$ to emphasize the fact that the mass field is for Lagrangian space, and recall that the $\langle\delta_L^k\rangle$ terms are inserted to guarantee that $\langle\delta_b\rangle = 0$.  As an explicit example, we will consider the Lognormal transformation, which aims to map $1+\delta_b \to \exp(b\delta_L)$.  The parameter $b$ is the one free parameter of this transformation:  large values of $b$ strongly enhance large values of $\delta_L$.

Since the coefficients in the Taylor series are simply $b_k = b^k$, for a Gaussian distribution of $\delta_L$, the additive correction terms can be summed explicitly:
\begin{equation}
 \sum_{k>0} \frac{b_k}{k!}\,\langle\delta_L^k\rangle
  = \sum_{k>0} \frac{b^k}{k!}\,\langle\delta_L^k\rangle
  = \exp(b^2\langle\delta_L^2\rangle/2) - 1.
\end{equation}
This means that, in this expansion, the Lognormal model is really 
\begin{equation}
 \delta_{b} = \exp(b\delta_L) - 1 - [\exp(b^2\,\langle\delta_L^2\rangle/2)-1]
            = \exp(b\delta_L) - \exp(b^2\,\langle\delta_L^2\rangle/2).
\end{equation}
Notice that, as a result of this additive normalization term, 
$\delta_b\ne 0$ when $\delta_L=0$.  

The cross-correlation between the biased field and the original one 
can also be done analytically:  
\begin{equation}
 \langle\delta_b\delta_L\rangle
  = \sum_{k>0} \frac{b_k}{k!} \,\langle\delta_L^{k+1}\rangle
  = b \,\langle\delta_L^2\rangle \,
      \exp\left(\frac{b^2\langle\delta_L^2\rangle}{2}\right)
 \label{fgcross}
\end{equation}
where the final expression shows the result of evaluating the sum for the Lognormal coefficients.  
Finally, the auto-correlation, which can also be done analytically, yields:  
\begin{equation}
 \langle\delta_b^2\rangle
 = \langle[\exp(b\delta) - \exp(b^2\langle\delta_L^2\rangle/2)]^2\rangle
 = \exp(b^2\,\langle\delta_L^2\rangle) [\exp(b^2\,\langle\delta_L^2\rangle) - 1].
 \label{fgdbdb}
\end{equation}

\subsection{Additive versus multiplicative normalization of the Lognormal mapping}
   
The treatment above ensures that $\langle\delta_b\rangle = 0$ by making an additive correction.  However, because this additive correction term has, in effect, shifted the mean value of the transformation, one should really account for the fact that the definition of the mean density, with respect to which one would like to define the biased fluctuation field, has also been modified.  This modification corresponds to accounting for the fact that, prior to adding these terms, $\langle 1 + \delta_b\rangle \ne 1$.  

If one enforces 
 $\langle 1 + \delta_b\rangle = 1$ 
by using a multiplicative factor rather than by an additive one, then one would write the Lognormal transformation as
\begin{equation}
 1 + \delta_b = \exp(b\delta_L)\, \exp(-b^2\langle \delta^2_L\rangle/2).
 \label{rhon}
\end{equation}
The first term on the right hand side is the same deterministic transformation of the variable as before, and the second, $\exp(-b^2\langle\delta^2_L\rangle/2)$, is the multiplicative factor that is required to ensure that $\langle 1 + \delta_b\rangle = 1$.  Note that this factor depends explicitly on the scale $L$ on which the Lognormal transformation occurs, via $\langle\delta^2_L\rangle$.  On large scales, $\langle\delta^2_L\rangle\ll 1$, making the overall normalization factor $\to 1$.  In this limit, the Taylor series expansion of eq.~(\ref{rhon}) has coefficients $b_k = b^k$, but otherwise, these coefficients generally pick up $\langle\delta^2_L\rangle$ dependent multiplicative correction factors.  

Integrating over the underlying Gaussian distribution of $\delta_L$ shows that the cross-correlation between the biased field and the original one is
\begin{eqnarray}
 \langle(1+\delta_b)(1+\delta_L)\rangle =  1 + \langle\delta_b\delta_L\rangle
                                        = 1 + b\,\langle\delta_L^2\rangle ,
 \label{xicrossL}
\end{eqnarray}
whereas 
\begin{eqnarray}
 \langle(1+\delta_b)^2\rangle = \exp[b^2\,\langle\delta_L^2\rangle] .
 \label{xibb}
\end{eqnarray}
These expressions differ from eqs.~(\ref{fgcross}) and~(\ref{fgdbdb}) for $\langle\delta_b\delta_L\rangle$ and $\langle\delta_b^2\rangle$ by one and two powers, respectively, of the multiplicative normalization factor associated with the Lognormal transformation (see discussion following eq.~(\ref{rhon})).  Eq.~(\ref{xibb}) is the usual expression for the relation between the correlation function of the Lognormal field and that of the underlying Gaussian \cite{cj1991}, suggesting that the present analysis is correct, and that of the previous section is not.  I.e., normalizing $\delta_b$ by subtracting a constant, rather than normalizing $1+\delta_b$ by a multiplicative constant is ill-advised.  Moreover, notice that now the cross-correlation function is particularly simple:  
 $\langle\delta_b\delta_L\rangle = b\,\langle\delta_L^2\rangle$, 
with {\em no} higher-order terms (eq.~(\ref{xicrossL})).  

For what follows, it is useful to write out slightly more general expressions for the cross- and auto-correlations.  The cross-correlation between the Lognormally biased field and the original Gaussian one smoothed on a different scale $L'$ than that on which the bias was defined, and separated by a distance $r$, is 
\begin{equation}
 \langle \delta_b\delta_{L'}|r\rangle = b\,\langle\delta_L\delta_{L'}|r\rangle 
                                   = b\,\xi_{LL'}(r) ,
\end{equation}
where $\xi_{LL'}(r)$ denotes the correlation between the initial field when smoothed on scale $L$ and when smoothed on scale $L'$ and displaced by $r$.  We note again that this expression is exact -- it is remarkable that the cross-correlation function is just a linearly biased version of that of the underlying field, despite the fact that the transformation itself was highly nonlinear.  Peaks in Gaussian fields exhibit this same simplicity \cite{dcss2010}.  Note in particular that the linear bias factor is the first term in the Taylor series of $\delta_b$ in the $\langle\delta_L^2\rangle\to 0$ limit.  We show shortly that, although
$\langle \delta_b\delta_{L'}|r\rangle\propto \langle\delta_L\delta_{L'}|r\rangle$
is generic, the constant of proportionality is not necessarily $b_1$, the first term in the Taylor series of the mapping between $\delta_b$ and $\delta_L$.  

The cross-correlation between two differently biased tracers $b$ and $b'$, defined using transformations on scales $L$ and $L'$, and separated by $r$, can also be computed exactly:  
\begin{eqnarray}
 1 + \xi_{bb'}(r) \equiv \langle(1+\delta_b)(1+\delta_{b'})\rangle
 = \exp[bb'\,\xi_{LL'}(r)] .
 \label{xinm}
\end{eqnarray}
This can be expanded as a series in $\langle\delta_L\delta_{L'}|r\rangle$ to yield: 
\begin{equation}
 \xi_{bb'}(r) \approx bb'\langle\delta_L\delta_{L'}|r\rangle
    + (bb')^2\frac{\langle\delta_L\delta_{L'}|r\rangle^2}{2} + \ldots \,\, .
 \label{xinmapprox}
\end{equation}
Therefore, if $b=b'$ and $L=L'$ then $\xi_{bb}(r)\approx b^2\xi_{LL}(r)$, and 
\begin{equation}
 \frac{\langle \delta_b\delta_{L}|r\rangle^2}{\langle \delta_b\delta_b|r\rangle}
 \approx \xi_{LL}(r).
 \label{ratio2xi}
\end{equation}
It turns out that this is a generic feature of local Lagrangian bias; it holds even when the constant of proportionality between $\langle\delta_b\delta_{L}|r\rangle$ and $\langle\delta_{L}^2|r\rangle$ differs from $b_1$ of the Taylor series.  

Before moving on, we note that the ratio 
 $\rho^2\equiv \langle\delta_b\delta_L\rangle^2/\langle \delta_b^2\rangle\langle\delta_L^2\rangle$ 
is sometimes used to quantify the `stochasticity' of the bias, where $\rho^2 < 1$ is taken to imply stochasticity.  For the Lognormal mapping above, $\rho^2 = x/({\rm e}^x - 1)$ with $x\equiv b^2\langle\delta_L^2\rangle$.  This equals unity only when $\langle\delta_L^2\rangle=0$, despite the fact that the Lognormal mapping is explicitly deterministic.  Clearly, $\rho^2$ is not a good indicator of stochasticity.  

\subsection{Approximate transformation for peaks}\label{pks}
Suppose one smoothes a Gaussian random field with a filter of scale $L_p$.  When the peak point process is smoothed with a filter of scale $L\gg L_p$, then ref.~\cite{bbks1986} argue that it is useful to think of the smoothed field as defining a peak fluctuation field which is related to that of the initial Gaussian field by 
\begin{equation}
 1 + \delta_p = \exp(b \delta_L - c \delta^2_L/2)\,
                     \sqrt{1 + c\langle\delta_L^2\rangle}\,
                \exp\left(-\frac{b^2\langle\delta_L^2\rangle/2}
                                {1 + c\langle\delta_L^2\rangle}\right).
  \label{bpkapprox}
\end{equation}
(Note that we are normalizing using a multiplicative rather than additive factor.)  The free parameters $b$ and $c$ are related to the properties of the peak (e.g., its height and curvature); they determine a rather complex normalization factor that, as before, depends on the smoothing scale via $\langle\delta_L^2\rangle$ and tends to unity on large scales.  In this limit, the first few bias coefficients associated with this transformation are $b_1 = b$, $b_2 = (b^2 - c)$, $b_3 = (b^3 - 3 b c)$, and $b_4 = (b^4 - 6 b^2c + 3 c^2)$.  (The structure of these terms means that $b_k$ is very similar to the Hermite polynomial $H_k(b)$; in fact $b_k = H_k(b)$ when $c=1$.  This happens because the bias relation has the same form, $\exp(bt - t^2/2)$, as the generating function of the Hermites.)  Note that the quantity $b_3-b_1b_2 = -2bc \ne 0$, in contrast to the Lognormal studied earlier.  This will matter in section~\ref{renorm}.

For this relation, the cross-correlation between such peaks and the field at a different position when smoothed on a different scale is 
\begin{equation}
 \langle\delta_p\delta_{L'}|r\rangle 
 = \frac{b}{1 + c\langle\delta_L^2\rangle}\,\xi_{LL'}(r).
\end{equation}
As for the Lognormal transformation,  this expression is exact for the  peak approximated field  we considered, and it is linearly proportional to $\xi_{LL'}(r)$; there are no higher order terms.  However, in this case, the constant of proportionality equals the first term in the Taylor series of $\delta_p$ only in the $\langle\delta_L^2\rangle\to 0$ limit; in general, they are different.  

When the two smoothing scales are the same, then 
\begin{equation}
 1 + \langle\delta_p\delta_p|r\rangle = 
  \Bigl[1 - C^2\,\xi_{LL}^2(r)\Bigr]^{-1/2}\,
 \exp\left(\frac{B^2\xi_{LL}(r)}{1 + C\,\xi_{LL}(r)} \right),
\end{equation} 
where
\begin{equation}
\quad B\equiv \frac{b}{1 + c\langle\delta_L^2\rangle} 
 \quad {\rm and}\quad C \equiv \frac{c}{1 + c\langle\delta_L^2\rangle}.
 \label{xipk}
\end{equation}
To leading order in $\xi_{LL}$, this reduces to  
\begin{equation}
 \langle\delta_p\delta_p|r\rangle \approx
    B^2\xi_{LL}(r) + \frac{(B^2 - C)^2}{2}\,\xi_{LL}^2(r)
                 + \frac{(B^3 - 3BC)^2}{3!}\,\xi_{LL}^3(r) + \ldots
 \label{xipkapprox}
\end{equation}
I.e., to leading order this transformation also satisfies eq.~(\ref{ratio2xi}) even though this bias factor is not the same as the leading order term in the Taylor expansion.  On the other hand, if we view this final expression as the Taylor series expansion of the correlation function, then the coefficients of this expansion have the same structure as the $\sigma_{L}\to 0$ limit of $\delta_p$.  We return to this shortly.

\subsection{Normalizing density rather than overdensity: The general case}\label{general}
The previous sections showed that, for Lognormal-like, local, deterministic, Lagrangian bias functions, one should normalize the density of the biased field to unity (using a multiplicative factor), rather than the overdensity to zero (by subtracting a constant).  We now study the general case.  

If one starts with an arbitrary bias function of the form given by eq.~(\ref{lb}), then the correctly normalized bias field is simply defined by eq.~(\ref{lbcorrect}):
\begin{equation}
 \delta_{B} \equiv \frac{1+\delta_{b} - \langle 1+\delta_b\rangle}
                        {\langle 1+\delta_b\rangle}
 = \frac{\sum_{k=1}^{\infty} (b_k/k!) (\delta_L^k - \langle\delta_L^k\rangle)}
   {\sum_{k=0}^{\infty} (b_k/k!) \langle\delta_L^k\rangle}.
 \label{deltabcorrect}
\end{equation}
Note that this renders the coefficient $b_0$ redundant, so we can set it equal to unity (and redefine all other coefficients in units of $b_0$).  This expression only differs from eq.~(\ref{lb}) because of the term in the denominator.  In general, $\delta_B\ne 0$ when $\delta_L=0$.  

It is a simple matter to check that this works out correctly for a Lognormal.
Moreover, it is straightforward to see that 
\begin{equation}
 \langle\delta_{L'}\delta_{B}|r\rangle 
 = \frac{\sum_{k=1}^{\infty} (b_k/k!) \langle\delta_{L'}\delta_L^k|r\rangle}
   {\sum_{k=0}^{\infty} (b_k/k!) \langle\delta_L^k\rangle}
 = 
   \frac{\sum_{k=1}^{\infty} (b_k/k!) \langle\delta_L^{k+1}\rangle/
                                   \langle\delta_L^2\rangle}
   {\sum_{k=0}^{\infty} (b_k/k!) \langle\delta_L^k\rangle}\,\xi_{LL'}(r)
 \equiv \,B_L \, \xi_{LL'}(r),
 \label{Xcorrect}
\end{equation}
where the second equality used the fact  that
\begin{equation}
 \langle \delta_{L'}|\delta_L\rangle
 =\delta_L\, \frac{\langle\delta_{L'}\delta_L|r\rangle}{\langle\delta_L^2\rangle}
 = \delta_L \,\frac{\xi_{LL'}(r)}{\sigma^2_{LL}}
\end{equation}
 to write 
\begin{eqnarray}
 \langle \delta_{L'} \delta_{L}^k|r\rangle
  &=& \int{ d\delta_{L} \mathcal{P}\left(\delta_{L}\right) \delta_{L}^k }
       \int{d\delta_{L'} \mathcal{P}\left(\delta_{L'}|\delta_{L}\right)\delta_{L'}}
   =  \int{ d\delta_{L} \mathcal{P}\left(\delta_{L}\right) \delta_{L}^k 
       \langle \delta_{L'}|\delta_L\rangle} \nonumber\\
  &=& \int d\delta_L\, \mathcal{P}(\delta_L) \delta_L^k\, 
       \delta_L \,\frac{\xi_{LL'}(r)}{\langle\delta_L^2\rangle}
   = \frac{\langle\delta_L^2\rangle}{\langle\delta_L^{k+1}\rangle}\,\xi_{LL'}(r).
\end{eqnarray} 
Equation~(\ref{Xcorrect}) shows that $\langle\delta_{L'}\delta_{B}|r\rangle$ is linearly proportional to $\xi_{LL'}(r)$, where the final equality defines the constant of proportionality $B_L$.  Note that this is an exact statement, valid for any $r$, $L$ or $L'$, and for any local deterministic bias function.  In particular, there is no requirement that $r$ be large compared to either $L$ or $L'$, nor that $L'$ be larger than the scale $L$ on which the local transformation from the dark matter to the biased field is (assumed to be) monotonic and deterministic.  Furthermore, notice that, although $B_L\ne b_1$ in general, they are indeed equal in the limit $\langle\delta_L^2\rangle\to 0$.  (Our eq.~(\ref{Xcorrect}) is consistent with eq.~(10) of \cite{mss2010}, in the Lagrangian bias limit in which all their $C_{pq}=0$; but our formulation highlights the fact that, in their expansion, all the higher order terms are proportional to powers of $\langle\delta_L^2\rangle$, whereas none are proportional to powers of $\xi$.)

For similar reasons the auto-correlation function of the biased tracers will reduce to a series of the form 
\begin{equation}
 \langle\delta_{B'}\delta_{B}|r\rangle = B_L^2\,\xi_{LL}(r) 
                                       + \frac{C_L}{2}\,[\xi_{LL}(r)]^2 + \ldots
 \label{Acorrect}
\end{equation}
This means that, to lowest order, eq.~(\ref{ratio2xi}) is satisfied even in the general case.  Note that the auto-correlation function of the biased tracers can always be written as a series in $\xi$.  E.g., terms involving derivatives or convolutions of $\xi$ do not appear.  Of course, this means that the power spectrum of a locally biased tracer will generically involve convolutions of the original $P(k)$.  In this sense, local bias is simpler in real space than it is in Fourier space.

\subsection{Renormalized bias}\label{renorm}

The analysis above highlighted the fact that it was important to include the multiplicative normalization factor when enforcing $\langle\delta_b\rangle = 0$.  We illustrated this using the Lognormal, for which all the sums could be performed analytically.  The question now arises as to what to do when this cannot be done.  

The analysis above suggests that one should redefine the mean density, and hence all bias factors, order by order.  This corresponds to truncating eq.~(\ref{lbcorrect}) rather than eq.~(\ref{lb}):  
\begin{equation}
 \delta_{B}^{(j)} = \frac{1+\delta_{b} - \langle 1+\delta_b\rangle_j}
                        {\langle 1+\delta_b\rangle_j}
 = \frac{\sum_{k=1}^{j} (b_k/k!) (\delta_L^k - \langle\delta_L^k\rangle)}
   {\sum_{k=0}^{j} (b_k/k!) \langle\delta_L^k\rangle}.
 \label{deltabj}
\end{equation}
Hence, the cross-correlation between the mass and biased fields is 
\begin{equation}
 \langle \delta_{L'}\delta_B^{(j)}|r\rangle 
 = \frac{\sum_{k=1}^{j} (b_k /k!)\langle\delta_L^{k}\delta_{L'}|r\rangle}
         {\sum_{k=0}^{j} (b_k/k!)\langle\delta_L^k\rangle}.
 \label{xicrossj}
\end{equation}
To 4th order in $\delta_L$, this is 
\begin{equation}
\langle \delta_{L'}\delta_B^{(4)}|r\rangle 
 = \frac{b_1\,\langle\delta_{L'}\delta_L|r\rangle
         + (b_3/3!)\,\langle\delta_{L'}\delta_L^3|r\rangle}
     {1+ (b_2/2) \langle\delta_L^2\rangle + (b_4/4!)\langle\delta_L^4\rangle}
  = \left[b_1+\left(\frac{b_3}{2}-\frac{b_1b_2}{2}\right)
                    \langle\delta_L^2\rangle\right]\xi_{LL'}(r)
  \equiv b_\times^{(4)}\xi_{LL'}(r),
\end{equation}
where the final equality defines $b_\times^{(4)}$.  
This differs from truncating eq.~(\ref{lb}) because of the term which is proportional to $b_1b_2$.  For the Lognormal transformation ($b_k = b_1^k$), $b_3 = b_1b_2$ so the term that is higher order in $\sigma$ is zeroed out in this expansion, whereas it would survive if we had started from eq.~(\ref{lb}).  Thus, this approach correctly gets $\langle \delta_L\delta_B^{(4)} \rangle = b\,\xi(r)$.  In fact, one can show that, for the Lognormal,
 $\langle \delta_L\delta_B^{(j)} \rangle = b\langle\delta_L^2\rangle$ for any $j$.  In this sense, this treatment is a significant improvement on the usual truncation of eq.~(\ref{lb}).  However, since $b_3 \ne b_1b_2$ in general, this approach would also lead one to conclude wrongly that the cross-correlation includes higher order terms when it does not.  For example, for the approximate peaks-bias relation of eq.~(\ref{bpkapprox}),
 $b_3 - b_1b_2 = -2bc \ne 0$.  
On the other hand, note that this expression returns 
 $b_\times^{(4)} \equiv b - 2bc\langle\delta_L^2\rangle/2 = b(1 - c\langle\delta_L^2\rangle)$ which is indeed the exact answer, 
$b/(1 + c\langle\delta_L^2\rangle)$,
expanded to first order in $\langle\delta_L^2\rangle$.  That is to say, although the renormalization approach suggests more complexity than is present in the exact answer, it is at least self-consistent.  

A similar calculation for the auto-correlation yields
\begin{equation}
 \langle \delta_B^{(4)}\delta_B^{(4)}|r\rangle \equiv 
 \xi_{bb}^{(4)}(r) = (b_\times^{(4)})^2\,\xi_{LL}(r) + \frac{b_2^2}{2}\,\xi_{LL}(r)^2.
\end{equation}
Note that although truncating eq.~(\ref{lb}) has the same form, in this case, the coefficients are correct for a Lognormal, illustrating again that this is a better approach.  

We are not the first to advocate normalizing by a multiplicative factor.  Ref.~\cite{pm2006} noted that this was advisable, especially in the context of truncated expansions.  However, that analysis did not highlight the fact that this makes the Lagrangian bias so simple (our eq.~ (\ref{Xcorrect})).  Rather, ref.~\cite{pm2006} went on to consider the implications in Fourier space, arguing that the $k=0$ limit of the term which scales as $b_2^2/2$ should be removed from the definition of the bias, and instead absorbed into a shot-noise like term. This procedure has the virtue of making the bias defined by the auto-correlation equal the square of that defined by the cross-correlation by definition even in Fourier space, which is the scaling satisfied by the exact answer.  Ref.~\cite{cs2012} argue that this may not be the best way to think about $b_2^2/2$ (or the higher order bias coefficients).  
 

\subsection{Relation to Szalay (1988)} \label{hermiteBias}
In all essential respects, the analysis above is simply a restatement of results in \cite{as1988}.  For $y\equiv\delta_L/\langle\delta_L^2\rangle^{1/2}$, Szalay assumed that the bias function $G(y)$ could have non-negative values only, and that 
\begin{equation}
 \langle G(y)\rangle = \int dy\,G(y)\,\frac{e^{-y^2/2}}{\sqrt{2\pi}} = 1.
\end{equation}
Because $G\ge 0$, and $\langle G\rangle=1$, his $G$ is, in effect, our $1+\delta_B$ of eq.~(\ref{deltabcorrect}).  He then expanded $G$ in terms of Hermite polynomials:
\begin{equation}
 G(y) = \sum^{\infty}_{k=0}\frac{B_k}{k!}\,H_k(y) \qquad {\rm where}\qquad
 B_k = \int dy\,\frac{e^{-y^2/2}}{\sqrt{2\pi}}\,G(y)\,H_k(y)
     = \langle G(y)\,H_k(y)\rangle
 \label{Jkszalay}
\end{equation}
and 
\begin{equation}\label{hermite}
 H_k(y) = e^{y^2/2}\,\left(-\frac{d}{dy}\right)^k e^{-y^2/2},
 \qquad {\rm with}\qquad \langle H_m(y)\,H_n(y)\rangle = \delta_{nm}\,m!.
\end{equation}
The orthogonality of the Hermite polynomials allowed him to show that 
\begin{eqnarray}
 \langle G(y_1)G(y_2)\rangle &=& 
  \int dy_1\, \frac{e^{-y_1^2/2}}{\sqrt{2\pi}}
 \int dy_2\, \frac{e^{-(y_2 - w_{12}y_1)^2/2(1-w_{12}^2)}}{\sqrt{2\pi(1-w_{12}^2)}}
    \sum^{\infty}_{k=0}\frac{B_k}{k!}H_k(y_1)
    \sum^{\infty}_{i=0}\frac{B_i}{i!}\,H_i(y_2) \nonumber\\
  &=& \sum_{k=0}^\infty \frac{B_k^2}{k!}\,w_{12}^k \, , 
 \label{szalayXi}
 \end{eqnarray}
where $w_{12} \equiv \langle y_1y_2|r\rangle \equiv\xi_{LL}(r)/\sigma_{\rm L}^2$.  Comparison with eq.~(\ref{Acorrect}) shows that his $B_k$ are our $B_L$, $C_L$, etc., so they are complicated sums over the (renormalized) bias factors.  This means they are, in general, $\sigma_{\rm L}$ dependent combinations of the coefficients $b_k$ of the Taylor series expansion of $\delta_h$ on scale $L$.  

Szalay did not compute the cross-correlation function, but it is straightforward to see that 
\begin{eqnarray}
 \Big\langle y_2\,G(y_1)\Big\rangle
  &=&  \int dy_1\, \frac{e^{-y_1^2/2}}{\sqrt{2\pi}}\,w_{12}\,y_1
       \sum^{\infty}_{k=0}\frac{B_k}{k!}H_k(y_1)
  = w_{12}\left\langle\sum^{\infty}_{k=0}\frac{B_k}{k!}H_k(y_1)H_1(y_1)\right\rangle \nonumber \\
  &=& w_{12}\, \Big\langle B_1\,H_1(y_1)^2\Big\rangle = B_1\,w_{12},
\end{eqnarray}
where we have used the fact that $H_1(y)=y$ and the orthogonality of the Hermite polynomials to simplify the expressions.  Notice that the tracer-mass cross-correlation is indeed just linearly proportional to the auto-correlation of the mass, and the square of it is the leading order term of the auto-correlation function, consistent with eq.~(\ref{ratio2xi}).  

We remarked earlier that the $B_k$ values depend on the scale on which the transformation is applied.  It is a simple matter to check that setting $B_k = (b\sigma_L)^k$ yields the correctly normalized Lognormal mapping considered earlier (eq.~(\ref{rhon})).  Notice that in this case $B_k$ is a separable function of a scale dependent piece $\sigma_L^k$, and a constant piece whose value is given by the large scale $\sigma_L\to 0$ limit of the Taylor series expansion of $\delta_h$.  That is to say, if one expands the (Lagrangian space) halo auto-correlation function in powers of the mass correlation $\xi_{LL}$, then the coefficient of the $k$th order term in the expansion is simply $B_k^2$, and this term is independent scale $L$, even though the bias coefficients in the Taylor expansion of the field $1+\delta_h(\delta_L)$ itself do depend $L$.  Therefore, the bias parameters estimated from a scatter plot of $1+\delta_h$ versus $\delta_L$ `run' with scale $L$, whereas those estimated from the correlation function do not.  

We will return to this in the next section, but note that it is not generic.  E.g., for the peaks transformation of eq.~(\ref{bpkapprox}), 
 $B_1 = B\,\sigma_L$, $B_2 = (B^2-C)\,\sigma_L^2$, 
 $B_3 = (B^3 - 3BC)\,\sigma_L^3$ etc., 
where $B$ and $C$ were defined in eq.~(\ref{xipk}).
I.e., the $B_k$ satisfy the same relations between the rescaled values $B$ and $C$ that the $b_k$ do for $b$ and $c$, but $B$ and $C$ differ from the peak-background split values (i.e. the large scale Taylor series coefficients of $\delta_h$) $b$ and $c$ by a factor of $1 + c\sigma_L^2$.

\subsection{Relation to halo bias from the excursion set approach}\label{halos}
The analysis above made the point that the leading order bias factor for the cross-correlation is the leading order term in the Taylor series expansion of $\delta_h$ only in the (large-scale) $\sigma_{\rm L}\ll 1$ limit.  So it is somewhat surprising that the differences between these two are sufficiently small as to have not attracted significant attention.  In part, this is because the large scale halo bias factors, as determined from the excursion set approach, satisfy 
\begin{equation}
 b_k = \nu^{k-1}\,H_{k+1}(\nu)/\delta_c^{k},\qquad{\rm where}\qquad \nu\equiv \delta_c/\sigma_h
 \label{bkhalos}
\end{equation}
\cite{mjw1997}.  Here $\sigma_h$ is related to the smoothing scale $R_h$ which contains the halo mass $M_h$ (i.e. $\bar\rho\, (4\pi/3)\,R_h^3 = M_h$), and $\delta_c$ is the overdensity associated with halo formation.  Since these bias factors were determined from a physically motivated model, rather than an arbitrary formal expansion, they are already correctly normalized, in the sense that $\langle 1 + \delta_h\rangle = 1$.  This makes the renormalized $\delta_B$ of eq.~(\ref{deltabcorrect}) equal to the original $\delta_h$, so the renormalized large scale bias factors are simply those in eq.~(\ref{bkhalos}). 

What is remarkable about the excursion set approach is that, for any smoothing scale (larger than that on which the halos were defined) the associated $B_k$ satisfy 
\begin{equation}
 B_k = \sigma_{\rm L}^k\, b_k 
 \label{Jkhalos}
\end{equation}
\cite{dcss2010, mps2012}.  I.e., just as for the Lognormal mapping (for which $b_k=b^k$), the $B_k$ are separable functions of the scale independent piece $b_k$ and the scale dependent piece $\sigma_{\rm L}$.  Note that this separability is not general; e.g., it does not apply for the mapping in eq.~(\ref{bpkapprox}). 
When $k=1$, this separability implies that $\langle\delta_h\delta_{\rm L}|r\rangle = \sigma_{\rm L}\,B_1\,w(r) = b_1\sigma_{\rm L}^2w(r) = b_1\,\xi(r)$.  I.e., for excursion set halos, the cross-correlation measurement returns the large scale linear bias factor $b_1$ whatever the smoothing scale $L$, and whatever the separation $r$.  

Inserting eq.~(\ref{Jkhalos}) in eq.~(\ref{szalayXi}) indicates that if one expanded the halo auto-correlation function in powers of the mass correlation, then the coefficient of the $k$th order term in the expansion is simply $b_k^2$.  Note again that this statement is not restricted to large scales.  This simple prediction for halos in Lagrangian space has not been noticed before.

In addition, the higher order correlations of the halos are given by Szalay's eq.~(11).  Like the two-point function, these can be written as sums of products of $\xi$.  To order $\xi^3$, the three point function is given by 
\begin{eqnarray}
 \zeta_{123} =&& b_1^2b_2\, (\xi_{12}\xi_{23} + \xi_{23}\xi_{13} + \xi_{12}\xi_{13})
              + b_2^3\,\xi_{12}\xi_{23}\xi_{13}\nonumber \\
              &&+ b_1b_2b_3\,\left[\xi_{12}^2\,\frac{\xi_{23} + \xi_{13}}{2} 
                + \xi_{23}^2\,\frac{\xi_{12} + \xi_{13}}{2}
                + \xi_{13}^2\,\frac{\xi_{12} + \xi_{23}}{2}\right],
\end{eqnarray}
where $\xi_{ij}$ denotes the two-point correlation function of the mass, smoothed on scale $L$ at separation $r_{ij}$.  For equilateral triangles, this simplifies to 
\begin{equation}
 \zeta_{\rm eq}(r) = 3\,b_1^2b_2\,\xi_{LL}^2(r) + b_2^3\,\xi_{LL}^3(r)
                   + 3\,b_1b_2b_3\,\xi_{LL}^3(r) .
\end{equation}

\subsection{On the appropriate ensemble over which to average}\label{porq}
The analysis above, like essentially all previous analyses in the literature to date, makes a technical assumption about how to compute the ensemble averages in Lagrangian space.  Namely, it assumes that, for Gaussian initial conditions, this average is over a Gaussian probability distribution function.  However, although averaging over the initial (Gaussian) pdf is technically correct for peaks, it is known to be incorrect for patches which are destined to form halos \cite{sl1999}.  This is because, in the excursion set definition of halos, $\delta_L$ is required to be less than $\delta_c$ on all scales $L$ larger than that on which the halo was defined:  this constraint on all larger scales is, in effect, a nonlocal requirement.  In practice, this means that the use of a Gaussian distribution for $\delta_L$ is, formally, inappropriate.  As shown in ref.~\cite{sl1999} (see their eq. (17)), a more careful treatment of this averaging, with the appropriate replacement of the Gaussian pdf yields
\begin{equation}
 \delta_c\,\frac{\bar\xi_\times}{\sigma_L^2} = H_2(\nu)
               + (\nu_{10}^2 + 1)\,{\rm erfc}{(\nu_{10}/\sqrt{2})} 
               - \sqrt{2\nu_{10}^2/\pi}\,e^{-\nu_{10}^2/2},
 \qquad{\rm where}\qquad 
 \nu_{10}^2 = \nu^2\, (\sigma^2_h/\sigma^2_L - 1).
 \label{b1q}
\end{equation}
When $\sigma_L\ll \sigma_h$, this expression asymptotes to $\bar\xi_\times \to b_1\,\sigma_L^2$ (recall that $b_1 = H_2(\nu)/\delta_c$), but in general the simple constant linear bias of the cross-correlation function between halos and mass is spoilt.  E.g., as $\sigma_L\to \sigma_h$, the cross-correlation $\bar\xi_\times$ tends to $\delta_c$.  

It is a simple matter to extend this analysis to see how averages over the higher order Hermite polynomials associated with the higher order bias factors are modified, but this is beyond the scope of the present work.  Measurements in simulations are needed to see if the range of scales over which the simple Gaussian averaged estimate is accurate -- and hence all the power of the Hermite polynomials -- is large enough to be interesting.

\section{Discussion}
In local bias models it is assumed that the biased field $1+\delta_b$ can be written as a function of the underlying density field at the same position.  We showed that, if the underlying field is Gaussian, then the cross-correlation between the biased field and the original one is linearly proportional to the auto-correlation function of the original field.  This is an exact result, valid on all scales, and is not a consequence of truncating expansions, etc., as most previous treatments assume.  While this is implicit in some previous work (e.g.~ref.~\cite{mss2010}), it has not been highlighted before.  

If one has been careful to ensure that $\langle 1+\delta_b\rangle = 1$ (by using eq.~(\ref{lbcorrect}) rather than eq.~(\ref{lb})), then the constant of proportionality is easily related to the first coefficient of the Taylor series expansion of the biased field, although they are not equal in general (eq.~(\ref{Xcorrect})).  In addition, we showed that, to leading order, the ratio of the square of the cross-correlation to the auto-correlation of the bias tracers equals the correlation function of the underlying field (eqs.~(\ref{Acorrect}) and~(\ref{ratio2xi})).  

We also explored the consequences of truncating these expansions, demonstrating that it is better to renormalize all truncated expansions (eq.~(\ref{deltabj})) than to not.  In this respect, our results agree with ref.~\cite{pm2006}.  Indeed, although our work has concentrated on Lagrangian bias, the multiplicative normalization (our eq.~(\ref{lbcorrect})) also holds also for Eulerian mass field, although providing an explicit expression when the mass field is not gaussian is more complicated (see ref.~\cite{pm2006} for the implications in Fourier space).  

Expanding the biased field using Hermite polynomials rather than powers of the overdensity, as was done by ref.~\cite{as1988}, provided a easy way to see a number of our results for local bias in the initial, Lagrangian, Gaussian field.  Although the coefficients $B_k$ of this expansion are functions of the scale on which the bias transformation is applied (eq.~(\ref{Jkszalay})), for the Lognormal distribution, as well as for excursion set halos, the scale dependence of these coefficients is trivial:  $B_k = b_k\sigma_{\rm L}^k$ (eq.~(\ref{Jkhalos})), where $b_k$ are the scale-independent peak-background split bias factors.  (The Lognormal has $b_k=b^k$, whereas excursion set halos have $b_k$ given by eq.~(\ref{bkhalos}).)  Therefore, if one expands the (Lagrangian space) halo auto-correlation function in powers of the mass correlation, then the coefficient of the $k$th order term in the expansion is simply $b_k^2$ (eq.~(\ref{szalayXi})).  These coefficients for the expansion of $1 + \xi_{hh}$ in powers of $\xi_{LL}$ are independent of scale $L$, even though the bias coefficients in the Taylor expansion of the field $1+\delta_h(\delta_L)$ itself do `run' with $L$.  This property of halo bias in Lagrangian space has not been emphasized before.  

The `running' of the $1+\delta_h(\delta_L)$ bias factors is most easily understood by noting that the Hermite polynomials arise naturally if one phrases the question of local bias in Fourier rather than real space \cite{tm2011}, a connection made recently in ref.~\cite[in particular, their Appendix~B]{mps2012}.  Moreover, ref.~\cite{tm2011} notes that it is better to work with `renormalized' parameters $c_n$ rather than the original bias parameters $b_n$ (but see \cite{mps2012} for how this renormalization should actually be done).  In effect, this corresponds to working with our eq.~(\ref{lbcorrect}) rather than eq.~(\ref{lb}); our analysis shows why this is necessary.  This connection to Fourier space bias is rich, because we showed that the simple relation $B_k = b_k\sigma_{\rm L}^k$ is not generic.  For the peaks transformation of eq.~(\ref{bpkapprox}), both $b_k$ and $B_k$ `run' with $L$.  Nevertheless, the $B_k$ satisfy the same relations between the rescaled values that the original $b_k$ (in the peak background split limit) do.  Exploring whether this is generic is the subject of ongoing work.  

Unfortunately, although the analysis based on Hermite polynomials is formally correct (and elegant), there are two reasons why, at least for describing halos, it cannot be valid for all smoothing scales $L$ and separations $r$.  First, halos do not overlap; this makes $\xi_{hh}$ of the halo point process tend to $-1$ on scales smaller than $\sim R_h$ \cite{mw1996}.  As a result, this halo exclusion limits the range in $r$ over which the local model can be applied.  The second is that, at least in the excursion set definition of halos, there is a nonlocal requirement on the density field:  this was the subject of section~\ref{porq}, which argued that this modifies the pdf over which one should compute ensemble averages.  Accounting for this makes the ratio of the halo-mass cross-correlation bias scale dependent (our eq.~(\ref{b1q}), which is eq. (17) of \cite{sl1999}).  The small scale limiting value of this modified expression yields $\delta_c$, which is the correct `one-halo' contribution to the Lagrangian space cross-correlation (the Lagrangian overdensity within a region which is destined to form a halo equals $\delta_c$ by definition), a fact which has not been emphasized before.  This suggests that averaging over the more appropriate pdf not only leads to sensible results, but accounts for halo exclusion as well, so it may be worth exploring further.  This may be particularly interesting because, in the large separation limit, it leads to $\xi_{hh} - b_\times^2\,\xi_{mm} < 0$ (eq.~(27) of \cite{sl1999}).  

There is a third reason why, at least for describing halos, the local bias model is unlikely to be valid for all smoothing scales $L$ and separations $r$.  This is related to the fact that the tidal field influences halo formation \cite{smt2001}.  The correlation with the tidal field leads, generically, to nonlocal bias even in Lagrangian space.  This is explored further in ref.~\cite{scs2012}, where the nonlocal bias terms are shown to matter most for massive halos, though the nonlocal effects are subdominant on large scales.  

If the gravitationally evolved nonlinear Eulerian mass field was a locally biased version of the initial field (it is not), then expanding in Hermite polynomials would also be the prefered way of describing halo bias (assuming the initial conditions were Gaussian).  This is because local Lagrangian bias, with local nonlinear evolution, leads to local bias with respect to the Eulerian field.  Conversely local bias with respect to the Eulerian field could be mapped back to local Lagrangian bias (with different bias factors, determined by the local linear-nonlinear mapping).  Some of the scalings which are characteristic of local Lagrangian bias will survive in Eulerian bias as well.  For example, the Lognormal transformation has one free parameter $b$.  If we use $b'$ to model the nonlinear mass field, and another $b$ to model the biased tracers (incidentally, this means that one should think of the biased field $1+\delta_b$ as being the nonlinear field $1+\delta_{b'}$ raised to the $b/b'$ power), then we may interpret our eq.~(\ref{xinm}) as the Eulerian space cross-correlation between the biased field and the nonlinear mass field.  Since $\langle\delta_b\delta_L|r\rangle$ and $\langle\delta_{b'}\delta_L|r\rangle$ equal $b\,\xi_{LL}(r)$ and $b'\,\xi_{LL}(r)$ for all $r$, one might naively have thought that $\langle\delta_b\delta_{b'}|r\rangle$ would also be linearly proportional to $\xi_{LL}(r)$.  Not only is this not true, eq.~(\ref{xinm}) shows that it is not proportional to $\xi_{b'b'}(r)$ either.  
In particular, eq.~(\ref{xinm}) shows that in this model of local Eulerian bias, the cross-correlation is linearly proportional to the auto-correlation of the mass only when $\xi_{LL}\ll 1$ (i.e., for sufficiently large $r$).   See Ref.~\cite{sw1998} for more discussion of this limit of local Eulerian bias.

Recent work has emphasized the fact that because nonlinear evolution is nonlocal, local Lagrangian bias will lead to nonlocal Eulerian bias, and vice versa \cite{tm2011,psp2012,css2012,bsdm2012}.  The Hermite polynomials are the orthogonal polynomials associated with a Gaussian field.  Therefore, the usefulness of Szalay's work for local Lagrangian bias suggests that, if bias is local with respect to the nonlinear, non-Gaussian field, then it would be natural to write Eulerian bias using the orthogonal polynomials of this non-Gaussian field.  This is the subject of work in progress.  

Finally, we note that our results depend only on the assumption that the Lagrangian matter field is Gaussian, so one might have thought that they also hold for modified gravity models.  However, for such models, $k$-dependence of the linear growth factor is generic, with the consequence that the assumption that Lagrangian halo bias is local is no longer so attractive \cite{phs2011}.

\acknowledgments
Thanks to M. Musso, A. Paranjape and R. Scoccimarro for helpful discussions.  
This work was supported in part by NSF 0908241 and NASA NNX11A125G.

\end{document}